\documentclass[useAMS,usenatbib]{mn2e}

\usepackage{subfigure}
\usepackage{footnote}
\usepackage{ulem}
\usepackage{color}
\usepackage{graphicx}
\usepackage{times} 
\usepackage{amssymb}
\usepackage{amsmath}
\usepackage{lscape}
\usepackage{url}
\usepackage{multirow}
\usepackage{longtable}
\usepackage[english]{babel}
\usepackage{float}

\newif\ifAMStwofonts
\AMStwofontstrue

\title[Abell 2146 Dynamics] {Dynamical analysis of galaxy cluster merger Abell 2146} \author[J.A. White et al.] {\parbox[]{6.in}
  { J.\,A. White$^{1,2}$\thanks{E-mail:
      jawhite@astro.ubc.ca}, R.\,E.\,A. Canning$^{3,4}$, L.\,J. King$^{1}$, B.\,E.\, Lee$^{1}$, H.\,R.~Russell$^{5}$, S.\,A~Baum$^{6,7}$, D.\,I.~Clowe$^{8}$, J.\,E.~Coleman$^{1}$, M.~Donahue$^{9}$, A.\,C.~Edge$^{10}$, A.\,C.~Fabian$^{5}$, R.\,M.~Johnstone$^{5}$, B.\,R.~McNamara$^{11}$, C.\,P.~O'Dea$^{6,7}$, J.\,S.~Sanders$^{12}$\\ } \\
  \footnotesize
  $^{1}$Department of Physics, University of Texas at Dallas, 800 W. Campbell Road, Richardson, TX 75080, USA \\
  $^{2}$Physics and Astronomy Department, University of British Columbia, 6224 Agricultural Road, Vancouver, BC V6T 1Z1, Canada\\
  $^{3}$Kavli Institute for Particle Astrophysics and Cosmology (KIPAC), Stanford University, 452 Lomita Mall, Stanford, CA 94305-4085, USA\\
  $^{4}$Department of Physics, Stanford University, 452 Lomita Mall, Stanford, CA 94305-4085, USA\\
  $^{5}$Institute of Astronomy, University of Cambridge, Madingley Road, Cambridge CB3 0HA, UK\\
  $^{6}$School of Physics \& Astronomy, Rochester Institute of Technology, Rochester, NY 14623, USA\\
  $^{7}$Department of Physics \& Astronomy, University of Manitoba, Winnipeg, MB R3T 2N2, Canada\\
  $^{8}$Department of Physics and Astronomy, Ohio University, 251B Clippinger Lab, Athens, OH 45701, USA\\
  $^{9}$Department of Physics \& Astronomy, Michigan State University, 567 Wilson Rd., East Lansing, MI 48824 USA\\
  $^{10}$Department of Physics, Durham University, Durham, DH1 3LE, UK\\
  $^{11}$University of Waterloo, Department of Physics \& Astronomy, Waterloo, Canada\\
  $^{12}$Max-Planck-Institut f\"ur extraterrestrische Physik, Giessenbachstrasse 1, 85748 Garching, Germany\\
}

\begin{document}

\date{...}

\pagerange{\pageref{firstpage}--\pageref{lastpage}} \pubyear{2014}

\maketitle

\label{firstpage}

\begin{abstract}
We present a dynamical analysis of the merging galaxy cluster system Abell 2146 using spectroscopy obtained with the Gemini Multi-Object Spectrograph on the {\it Gemini} North telescope. As revealed by the {\it Chandra} X-ray Observatory, the system is undergoing a major merger and has a gas structure indicative of a recent first core passage. The system presents two large shock fronts, making it unique amongst these rare systems. The hot gas structure indicates that the merger axis must be close to the plane of the sky and that the two merging clusters are relatively close in mass, from the observation of two shock fronts. Using 63 spectroscopically determined cluster members, we apply various statistical tests to establish the presence of two distinct massive structures. With the caveat that the system has recently undergone a major merger, the virial mass estimate is  $M_{\rm vir}= 8.5^{+4.3}_{-4.7} \times 10^{14} M_{\odot}$ for the whole system, consistent with the mass determination in a previous study using the Sunyaev-Zel'dovich signal. The newly calculated redshift for the system is $z=0.2323$. A two-body dynamical model gives an angle of $13^{\circ}-19^{\circ}$ between the merger axis and the plane of the sky, and a timescale after first core passage of $\approx$\,0.24-0.28 Gyr. 

\end{abstract}


\section{Introduction}
\label{intro}
Galaxy clusters are formed hierarchically through the mergers of smaller clusters and groups. Mergers of galaxy clusters are the most energetic events since the big bang, with a total kinetic energy that can reach 10$^{57}$\,J (e.g. \citealt{sar}), and major mergers close to the plane of the sky are very rare systems but of extreme importance in cosmology (e.g. \citealt{clowe, markevitch07}). Most of the mass in galaxy clusters is dark matter, and the bulk of baryonic mass is in the form of ionized plasma, comprising about 15\% of the total mass. During a merger, as the plasma clouds of each cluster pass through each other they are affected by ram pressure, causing them to slow down. The galaxies in a cluster are effectively collisionless and are affected primarily by tidal interactions during the merger.
Dark matter does not have a large cross-section for self-interaction (e.g. \citealt{markevitch04, randall}), so shortly after collision the plasma clouds are expected to lag behind the dark matter and the major concentrations of cluster galaxies (e.g. \citealt{clowe}). A merger thus results in an offset between the dominant luminous component, imaged using X-ray telescopes, and the total mass, mapped using gravitational lensing. 

\begin{figure*}
\centering
\includegraphics[width=\textwidth]{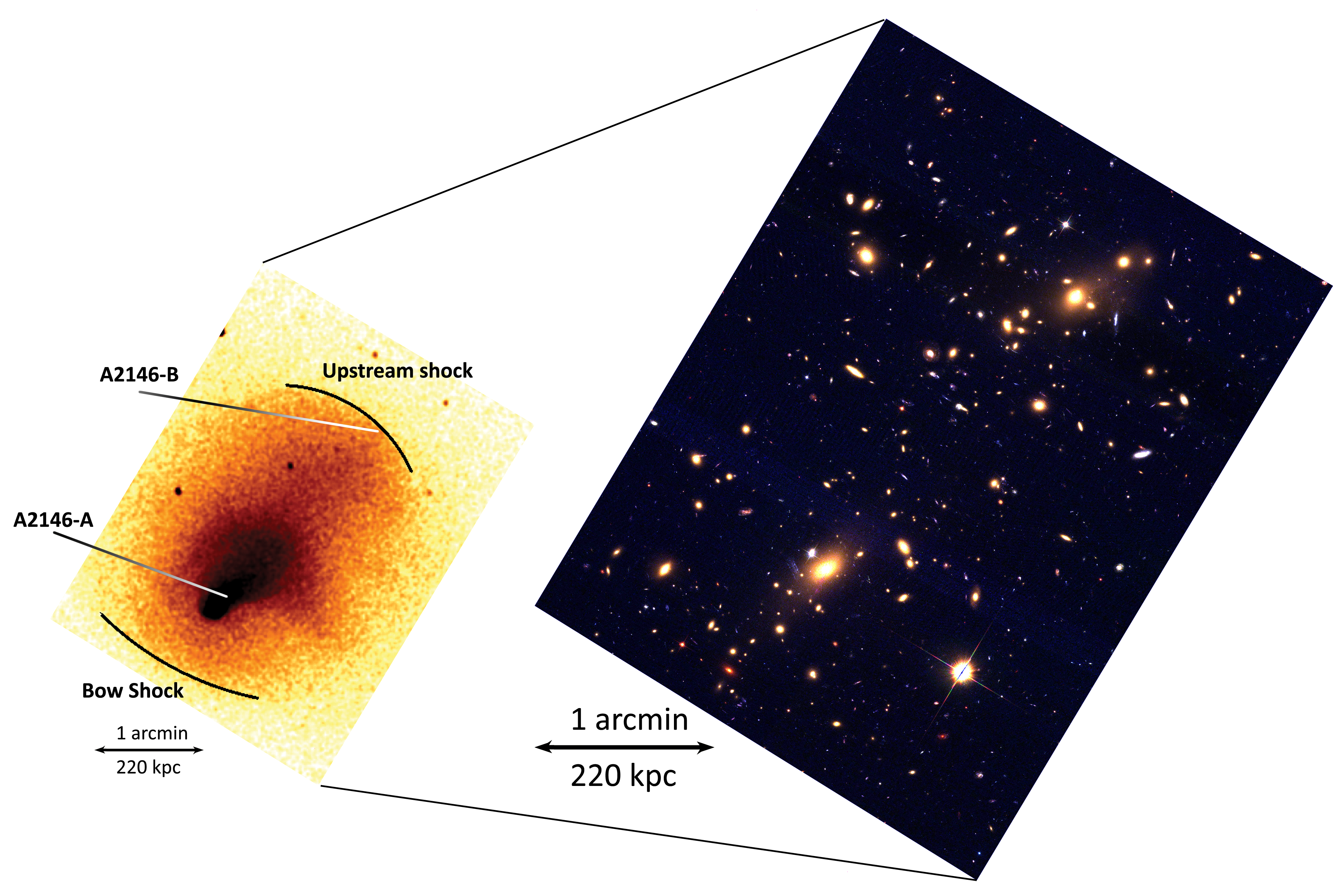} 
\caption{{\bf Left:} Exposure-corrected Chandra X-ray image for the 0.3-7 keV energy band (North is up and East is to the left).  The image has been smoothed with a 2D Gaussian of width 1.5\,arcsec.  {\bf Right:} HST F435W, F606W, F814W, colour composite image of Abell 2146. \label{xray}} 
\end{figure*}

Abell 2146 is in the throws of a major merger \citep{russell10} and X-ray maps show a structure similar to the ``Bullet Cluster" (\citealt{markevitch02, clowe}). Unique among merger systems, two clear shock fronts are detected in the {\it Chandra} X-ray temperature and density maps indicating Mach numbers of $\approx$\,2.3 for the bow shock and $\approx$\,1.6 for the upstream shock, and corresponding  velocities in the plane of the sky of 2700$^{+400}_{-300}$ km\,s$^{-1}$ and $2400\pm 300$ km\,s$^{-1}$ respectively \citep{russell12}. From the shock velocity and the projected distance from the collision site, \citet{russell10} estimate that the merger is observed $0.1-0.2$ Gyr after core passage. The detection of the shock fronts is fully consistent with the merger being observed soon after first core passage (e.g. \citealt{sar}), but the timescale is approximate due to uncertainty in the velocity of the subcluster - likely to be significantly lower than the shock velocity (e.g. \citealt{springel, milos07}) - and since the location of the collision site is not precisely known.

In Fig.\,1 we show the {\it Chandra} X-ray observatory image from \citet{russell12} and {\it Hubble Space Telescope} ({\it HST}) optical colour composite of Abell 2146 (Program 12871, PI King). Subclusters A and B are labeled on the left, with subcluster A being located at the head of the `bullet' in the south-east. The Brightest Cluster Galaxy (hereafter BCG) of the SE subcluster (A2146-A), is in an unexpected location lagging {\it behind} the X-ray cool core \citep{canning}. As noted above, in the early stages of a cluster merger the galaxies should be nearly collisionless and follow the dark matter distribution, which should lead the collisional X-ray gas (e.g. \citealt{clowe, markevitch}).

In addition to the strange projected location of the BCG in Abell 2146-A, the system is unusual in that despite it appearing to undergo a major galaxy cluster merger it does not have a detected radio halo \citep{russell11}. Studies of samples of radio halos in clusters have shown that they are found almost exclusively in morphologically disturbed clusters, indicating that their generation is associated with acceleration or re-acceleration of energetic particles in the highly shocked cluster gas, and that there is a relationship between the radio power of the halo, $P_{\mathrm{radio}}$, and the X-ray luminosity of the cluster, $L_{\mathrm{X}}$ \citep{liang, giovannini,buote, brunetti, rossetti, cassano10}. The conditions under which radio halos are generated in clusters, such as the cluster masses, velocities and timescales, are still open questions. Merging clusters, such as Abell 2146, which do not exhibit these features may be a powerful diagnostic to test models of radio halo generation.

\begin{figure*}
 \centering
\includegraphics[width=0.47\textwidth]{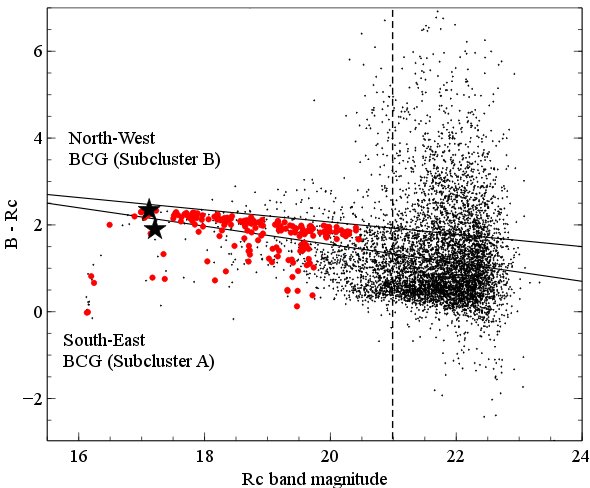}
\includegraphics[width=0.49\textwidth]{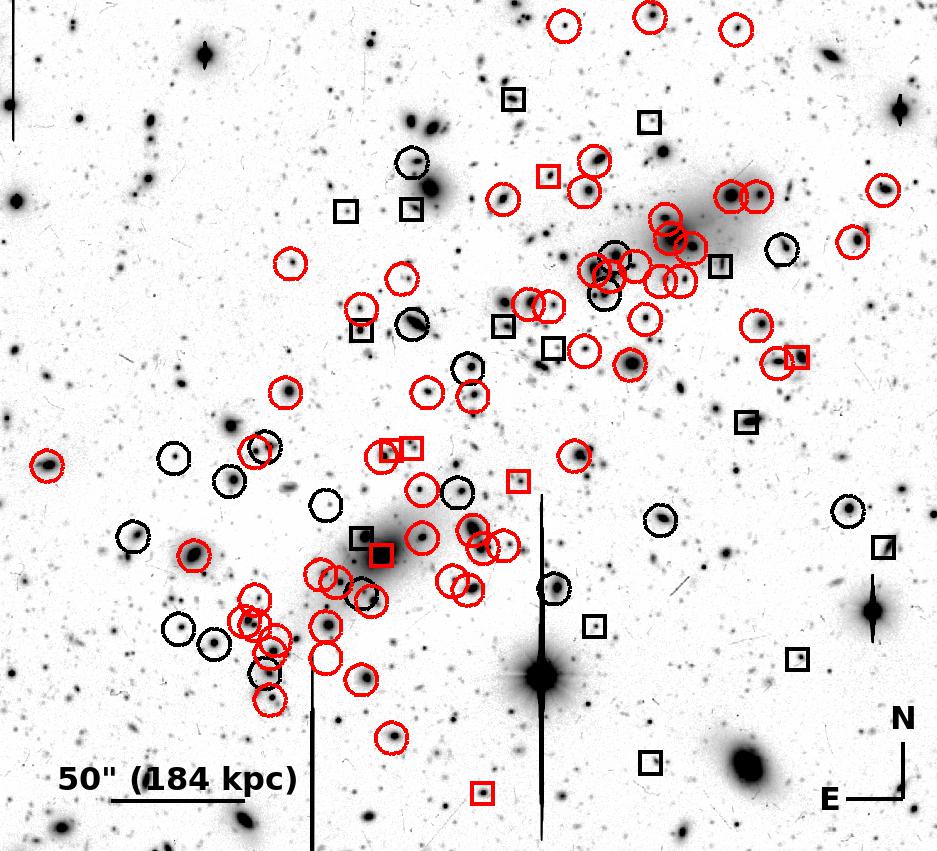}
 \caption{Target selection for GMOS spectroscopy: Targets were chosen based on colour, magnitude and cluster-centric distance, detected from our {\it Subaru} Suprime-Cam imaging.  {\bf Left:} Colour-magnitude diagram for galaxies within 3\,arcmin of the centre of the system.The dashed line indicates the initial apparent magnitude cut which we imposed on our sample. The red sequence was determined by fitting a Gaussian in magnitude bins of $18>\mathrm{Rc}$, $18<\mathrm{Rc}<19.5$, $19.5<\mathrm{Rc}<21$ and $21<\mathrm{Rc}$. The final 170 galaxies selected from our catalogue by the GEMINI mask making tools are indicated in red and the two BCGs are marked by black stars.  {\bf Right:} The 97 objects for which we determined redshifts overlaid on our {\it Subaru} Suprime-Cam composite colour image; those found to be cluster members are indicated in red and those not in the cluster indicated in white. The circles in each case indicate galaxies which fall on the red sequence while the boxes indicate galaxies blueward of the red sequence. \label{col_mag_image}}
\end{figure*}

In this paper we present multi-object spectroscopy of the system, in order to identify substructure and to provide a complementary method of probing the dynamical status, timescale of the merger, and mass of the system (e.g. \citealt{barrena, ferrari, girardi}).
Section 2 will briefly describe the optical sample selection, and observations and data reduction will be the subject of Section 3. Results from an analysis of the substructure within the Abell 2146 system will be presented in Section 4 and Section 5 will discuss these results in the context of simple dynamical models. Section 6 presents a summary and discussion of our results. A future paper will address the stellar populations of the system.

Throughout this paper we assume a $\Lambda$CDM cosmology with present day Hubble parameter H$_{0}=70\,$km\,s$^{-1}$\,Mpc$^{-1}$, matter density parameter $\Omega_{\rm m}=0.3$, and dark energy density parameter $\Omega_{\Lambda}=0.7$ (we assume a cosmological constant, equation of state parameter $w=-1$). For this cosmology and at the redshift of Abell 2146 ($z=0.2323$), an angular size of 1\,arcsec corresponds to a physical scale of $\approx 3.677$\,kpc.

\section{Optical Sample Selection}

Here we briefly present the criteria for selecting the galaxy sample for {\it GEMINI} GMOS-N multi-object spectrograph (MOS) observations of Abell 2146.  The next section will outline the observations and data reduction.

Targets were detected on our Rc band optical Subaru imaging using {\small SEXTRACTOR} \citep{sex}.
We restricted our target galaxies to those brighter than 21 apparent magnitude and to those within 3\,arcmin from the centre of the Abell 2146 system. We then determined the cluster ``red-sequence'' by binning the galaxies into magnitude bins of $18>\mathrm{Rc}$, $18<\mathrm{Rc}<19.5$, $19.5<\mathrm{Rc}<21$ and $21<\mathrm{Rc}$ mags. We fit two Gaussians to the bimodal population of colour in each magnitude bin; one defining the ``red sequence'' and one defining the ``blue cloud", and took the $3\sigma$  boundaries of the Gaussian defining the red sequence to determine its width in each magnitude bin. We then fit a linear relation to the red sequence (black solid lines in Fig. \ref{col_mag_image}). We identify 337 targets with Rc$<21$ mags on the cluster red sequence (see Fig. 2, left).

Many galaxies morphologically associated with the X-ray shock fronts and one of the BCGs do not lie on the red sequence. For this reason we additionally add 50 targets to our target list (including 12 acquisition targets) based on an inspection of the {\it Subaru} images. In particular we added the south-east BCG and targets close to the BCGs and X-ray shock front which were too blue to be included on the red sequence. We also added two possible strong lensing features to our target list. Finally we have a sample of 375 target objects and a further 12 acquisition objects to input to the  Gemini MOS Mask Preparation Software (GMMPS).

The two BCGs, galaxies close to the shock front and strong lensing features were assigned the highest priority. All other targets were prioritized by magnitude then cluster-centric distance and the GMMPS was iteratively run in order to fit as many objects as possible on the masks while requiring at least 3 acquisition objects per mask. Finally, we obtained 7 masks totaling spectra on 170 targets indicated by red points on the left hand panel of Fig. 2. Of the 170 targets observed with GEMINI 139 lie on the red-sequence giving a red sequence completeness of 67\% at Rc$<20.5$ mags within 3 arcmin of the center of the Abell 2146 system. As described in Section 3, we were able to obtain redshifts for 97 galaxies, of which 63 were determined to be cluster members and are tabulated in the Appendix (see Appendix and Section 4). The 97 objects are indicated on the right panel of Fig. 2 with the cluster members in red. Objects indicated by circles are those determined to be on the red sequence while those indicated by boxes are found blueward of the red sequence.  

\begin{figure*}
\includegraphics[width=0.48\textwidth]{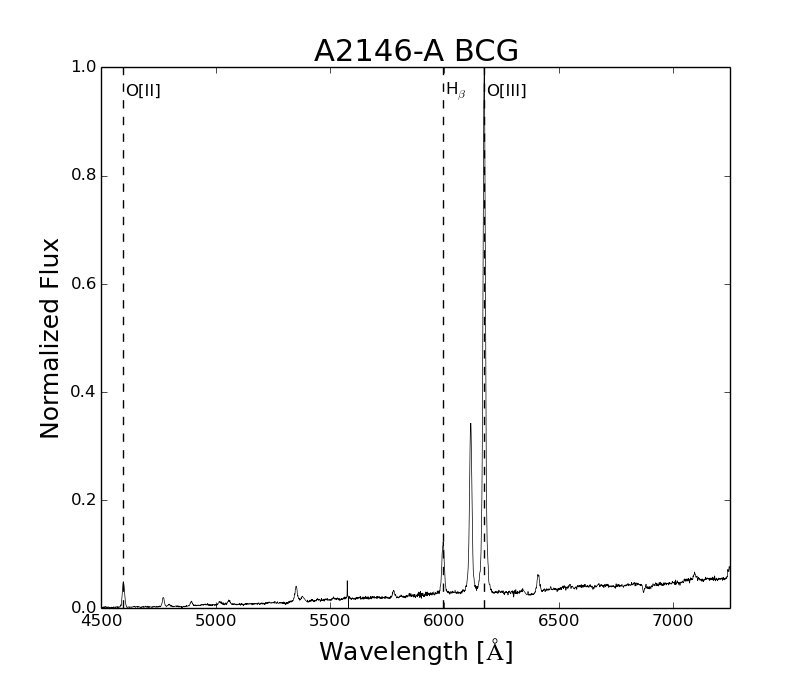}
\includegraphics[width=0.48\textwidth]{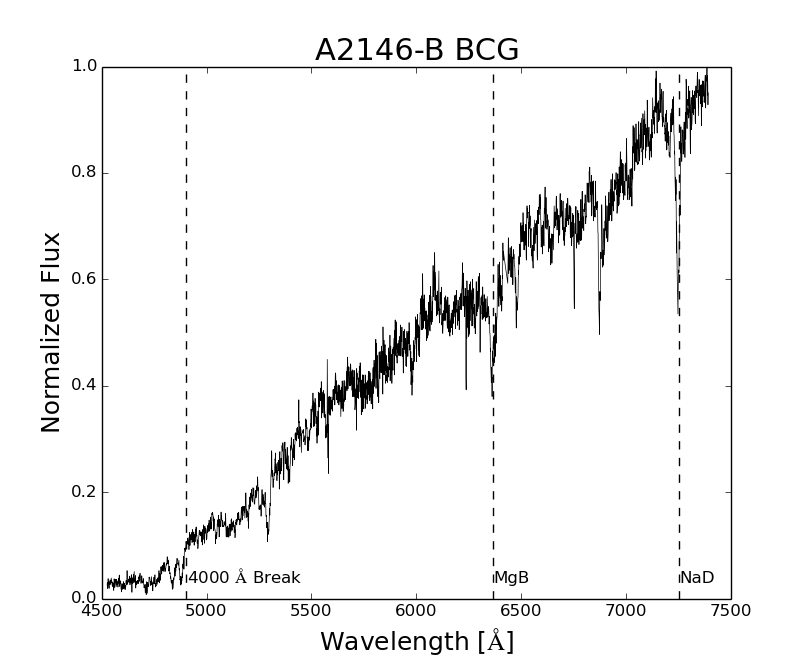}
 \caption{One dimensional extracted spectra of the BCGs in A2146-A (left) and A2146-B (right). Spectra are shown at the observed wavelengths. Various emission and absorption features are labelled. \label{BCG_b}}
\end{figure*}

\section{Observations and Data Reduction}
Spectroscopic observations of Abell 2146 were carried out with the {\it Gemini} Multi-object Spectrograph (GMOS) at the {\it Gemini North Telescope} on Mauna Kea, Hawaii. They were made over the course of four nights between 2012 April 19 and May 1 (Program ID: GN-2012A-Q-47, PI Canning). We used the GMOS-MOS instrument equipped with the B600+\_G5307 grating. This configuration yielded a dispersion of 0.50\AA/pixel. The 5.5\,arcmin field of view of the GMOS detector allowed observations to be made in a single pointing. 

Using slit widths of 1\,arcsec  and slit lengths of 5\,arcsec, between 18-31 objects were able to be targeted in each mask. Due to the densely packed nature of the field, 7 masks were used to obtain spectra of a total of 170 science targets. Three 1200 second exposures were taken with each mask in order to achieve a S/N$\sim3$ per spectral pixel, at the redshift of the 4000\AA\, break for the faintest galaxies in our target list (determined using GEMINI GMOS exposure time calculator). Wavelength calibration was performed using Helium-Argon lamps.

The data reduction was performed using the Gemini package in IRAF\footnote{IRAF is distributed by the National Optical Astronomy Observatories, which are operated by the Association of Universities for Research in Astronomy, Inc., under cooperative agreement with
the National Science Foundation.}; see Fig.\,3 for examples of the resulting spectra. All science exposures, comparison lamps, and flats were overscan/bias subtracted and trimmed. The resulting reduced spectra were then processed to remove cosmic rays, wavelength calibrated, extracted to 1D and combined.

The cross-correlation method was used to calculate radial velocities through the RVSAO\footnote{RVSAO was developed at the Smithsonian Astrophysical Observatory Telescope Data Center.} package in IRAF. A catalogue of 23 different templates from the RVSAO library \citep{rvsao}, including stars, quasars, and emission and absorption line galaxy template spectra was used to determine the radial velocities. In order to determine the templates to use in the final analysis we relied on a combination of visual examination and the {\it R}-value returned by RVSAO.
Each extracted 1D target spectrum was visually examined and a redshift identified by at least two members of the team, with a third examining spectra where disagreements were found. For 54/170 objects no redshift was able to be determined with a visual inspection due to a low S/N. For a further 19/170 targets the visual examinations yielded a redshift which was not consistent with the radial velocity measurements for any of the 23 RVSAO template spectra. In both these cases the targets were not used in the remainder of the analysis in this paper. In 73/170 targets the visual examination yielded a redshift within $\pm6,000$km~s$^{-1}$ of the RVSAO radial velocity with the highest {\it R}-value. In these cases the radial velocity corresponding to the highest {\it R}-value template was used. In the remainder (24/170), the highest {\it R}-value template did not correspond to the visually determined $cz$. However, a template with {\it R}-value smaller than the best value did have a radial velocity within $\pm5,000$~km~s$^{-1}$ of the visual velocity. If $\Delta~R<5$ then we accepted the fit with a lower {\it R} after further examination of the template spectra. In most of these 24 targets poor sky subtraction was the reason for poor template fitting. However, the HST imaging has shown many lensed features are apparent in the Abell 2146 system and galaxy density is high. Some poor redshifts may be due to blending between two of more sources. The spectra of the BCGs in A2146-A and A2146-B are shown in Fig.\,3.

In order to minimize errors associated with cross-correlation and to correct for artifacts in the spectra (such as cosmic rays), three separate observations of each object were made and then the spectra were co-added together (e.g. \citealt{gallazzi}). This signal averaging helps to increase the signal to noise of the spectrum and can help to remove certain undesirable features that are only contained in one of the individual spectra. The radial velocities and errors were then compared to both the co-added and individual spectra.

\section{A2146 Structure}

\subsection{Cluster Member Determination}

The target radial velocities were used to determine cluster membership: first, we made a $\pm 10,000$ km\,s$^{-1}$ cut at the previously calculated cluster redshift of $z=0.2343$ \citep{red}, as was done by \cite{ferrari}. Cluster members were then determined by a 2.5$\sigma$ clip of the remaining objects. This leaves a total of 63 galaxies out of the 97 objects for which we could determine a redshift. Fig.\,2 shows the 63 objects identified as cluster members circled in red. The foreground and background objects are identified in white. The velocity distribution for the 97 targets (which are not stars) is shown in the top panel of Fig. 4 while the bottom panel shows only the 63 cluster members after $\sigma$ clipping.

\begin{figure}
 \centering
\includegraphics[width=0.48\textwidth]{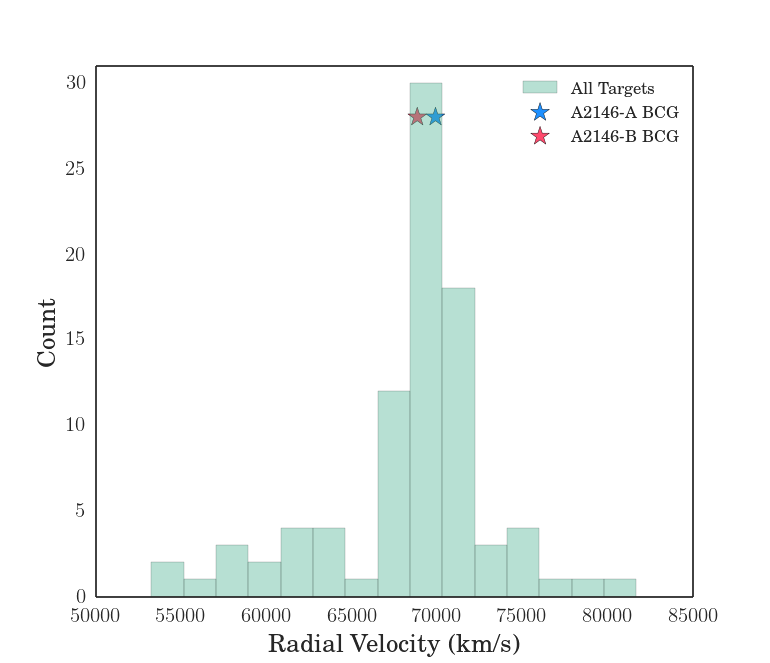}
\includegraphics[width=0.48\textwidth]{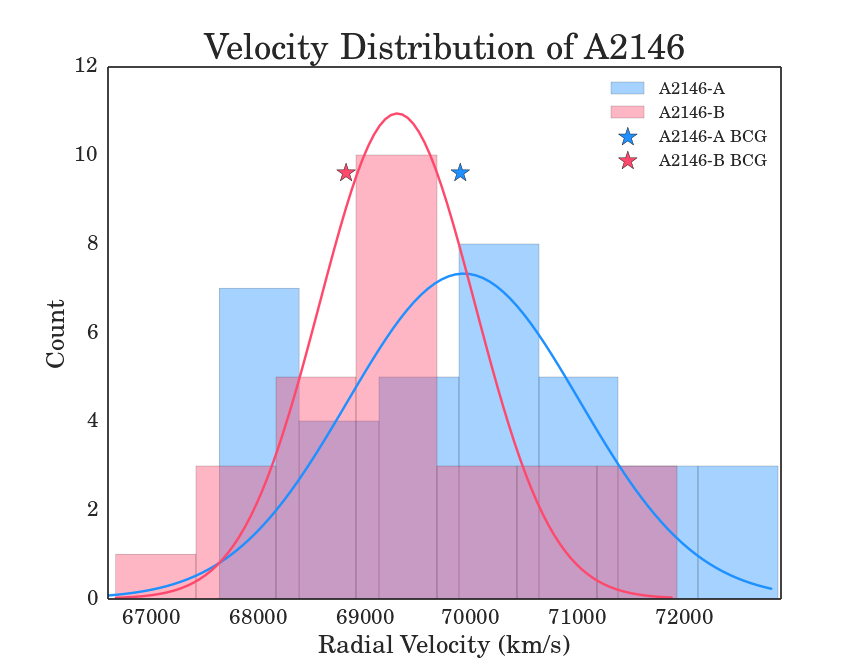}
 \caption{{\bf Top:} The radial velocity histogram for all objects where a reliable redshift was determined and the targets are not stars. The cluster radial velocities peak around 70,000~km~s$^{-1}$, however, there is no evidence in our data that the broad tails in the distribution correspond to further clustering of galaxies. {\bf Bottom:} Radial velocity distribution of cluster members. Cluster members in A2146-A are shown in blue, and A2146-B in red. The solid lines indicate a Gaussian profile at the biweight mean. The radial velocities of the two BCGs are denoted by stars. Subcluster membership was determined using the KMM algorithm \citep{ashman}. \label{red_dist}}
\end{figure}

\begin{figure}
\includegraphics[width=0.5\textwidth]{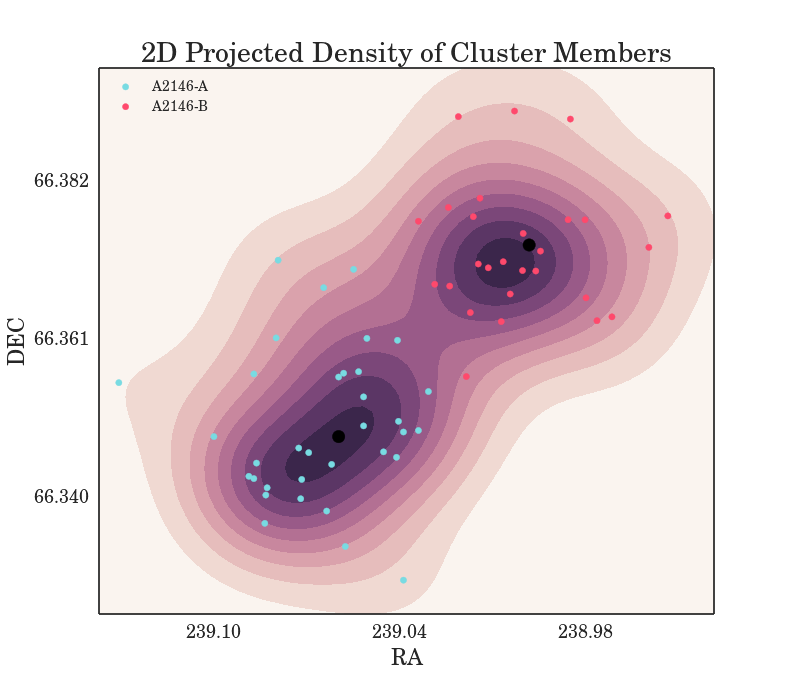}
 \caption{Projected density map of cluster members in 2D. The relative density contour map is shown in position space. The subcluster A2146-A is on the bottom left (blue circles) and A2146-B (red circles) is on the top right with the BCGs in both subclusters represented with a black point. In the figure North is up and East is to the left.  \label{density}}
\end{figure}

\subsection{Subclustering}
\label{subclustering}
The velocity distribution of the Abell 2146 system is shown in Fig.\,\ref{red_dist}. We apply a Lilliefors test to determine whether the velocity distribution of the 63 cluster member galaxies is statistically different from a Gaussian distribution. We calculate a p-value of 0.23 indicating that the null hypothesis, that the velocity distribution of the system as a whole is Gaussian distributed, cannot be rejected. However, for a plane of sky merger this is not a sufficient condition to exclude the presence of subclustering.

To investigate the possibility of subclustering further we apply the Dressler-Schectman test \citep{DS} (henceforth DS test) and also carry out an analysis of the 2D projected density of cluster members.

First, we apply the DS test which provides a useful statistic to determine the likelihood of subclustering. The DS test compares the kinematics of a sample of size $N_{nn}$ to the kinematics of the whole system. For each galaxy in the field, the distances are computed between every pair in the system and then the $N_{nn}$ nearest neighbours are chosen for analysis. The statistic is given by

\begin{equation}
\delta_{i}^{2} = \frac{N_{nn}}{\sigma_{v}^{2}}[(\bar{v}_{local} - \bar{v})^{2} + (\sigma_{local} - \sigma_{v})^{2}]\,,
\end{equation}
where $\sigma_{v}$ is the velocity dispersion of the whole group and $\sigma_{local}$ is the velocity dispersion of the $i^{th}$ sample. These values are summed to obtain

\begin{equation}
\Delta = \sum_{i}^{N} {\delta_{i}}\,,  
\end{equation}
with $\frac{\Delta}{N} > 1$ being an indicator of substructure present in the field. Using $N_{nn} = \sqrt{N_{Total}}$ we found a $\frac{\Delta}{N} = 1.08$. The marginal result makes it hard to determine if there is substructure in the system (see also \citet{pinkney} for the results of applying the DS test to simulated galaxy clusters).

To test if the calculated $\Delta$ was significantly different from random, we randomly shuffled the velocities of the objects while holding the positions fixed. Repeating this process 10,000 times, we found that the fraction of random velocities that gave a larger $\Delta$ was $P=0.18$, where a value of $P$ less than $0.1$ would be expected for a non-random distribution. Thus, we conclude that the DS test results are not strong enough to indicate substructure in the system. We return to this result in the Discussion.

Second, we analyse the 2D projected density of all of the cluster members. Cluster members identified by their redshift were plotted in RA and DEC space with overlaying contours determined using a Kernel Density Estimator (KDE) and a Gaussian kernel. The KDE utilized a Silverman bandwidth. Since Gaussian kernels are less sensitive to bimodal distributions, this test will be biased towards strong substructure detections. The results plotted in Fig.\,5 clearly show that two density peaks are present in the data. This confirms the existence of at least two subclusters and is consistent with the {\it Chandra} X-ray data which shows a bi-model merger morphology with both forward and reverse shocks detected in the X-ray analysis and along the same axis as our two subcluster clumps \citep{russell12}. The KDE analysis allows us to conclude that there is substructure present in the system.

Having confirmed the presence of substructure we wish to identify the cluster members of particular subclusters in order to characterise properties such as the mass of the substructures, the timescale since core passage of the merger and the presence of any offsets between the galaxy and hot gas populations. In order to assign the cluster members to particular subclusters we apply the KMM algorithm \citep{ashman}. This determines the mixture of multi-dimensional Gaussians which are best fit to a set of data and the probability of each data point belonging to each Gaussian.

The clear X-ray shock fronts and the small line-of-sight velocity offsets of the two BCGs indicates strongly that our merger is likely to be in the plane of the sky. As such, the distribution of velocities from each subcluster is expected to have a significant overlap (as seen in Fig.\ref{red_dist}). Including such similar velocity information in the KMM algorithm can lead to non-physical results in distinguishing between the two subcluster components. The analysis was carried out with position data only and yielded results which correspond well with the projected galaxy density peaks seen in Fig. \ref{density}. The KMM algorithm classified 35 galaxies as belonging to A2146-A and 28 belonging to A2146-B. These subcluster members are marked in Fig. \ref{density} (with blue corresponding to A2146-A and red corresponding to A2146-B), and subcluster membership is also noted in Table 1 in the appendix. The velocity distribution of the galaxies identified in the two subclusters is shown in Fig. \ref{red_dist}. The KMM algorithm is sensitive to outliers \citep{ashman} which is of particular importance for a cluster sample such as ours where the data is incomplete. We therefore bootstrap the sample (e.g. Waters et al. 2008) and re-run the KMM test 5,000 times to determine which galaxies to assign to which subcluster and the centroid of our two subclusters. The centroids and their 68 per cent uncertainties are 239.062$^{+0.004}_{-0.004}$, 66.349$^{+0.002}_{-0.002}$ (15:56:14.9$^{+0.9}_{-0.9}$, 66:20:56.4$^{+7.2}_{-7.2}$) and 239.005$^{+0.005}_{-0.005}$, 66.373$^{+0.002}_{-0.002}$ (15:56:1.2$^{+1.2}_{-1.2}$, 66:22:22.8$^{+7.2}_{-7.2}$) for subclusters A and B respectively.

Applying a Lilliefors test again to the individual subclusters yields a p-value of 0.83 for A2146-A and a p-value of 0.059 for A2146-B. This indicates that the velocity distribution for A2146-A can be adequately described by a Gaussian distribution. A Gaussian distribution for the velocities of subcluster A2146-B would be rejected at the 94 ($<2\sigma$) per cent level. There is evidence from a weak gravitational lensing mass reconstruction of {\it HST} data that A2146-B is rather more extended that A2146-A, possibly due to an intrinsically more elongated mass distribution (King et al. 2015). The numbers of galaxies for which we have spectroscopic data is insufficient to decide between these possibilities.

\subsection{Velocity Dispersions of the subclusters}
\label{velocity}

The biweight estimator  \citep{beers2} was applied to the cluster members for location and scale giving a mean apparent cluster redshift of $z=0.2323$ (radial velocity of $69,640$\,km\,s$^{-1}$). The velocity dispersions were calculated from the median absolute deviation (MAD) (e.g. \citealt{ferrari}) with errors estimated through a bootstrap technique. The velocity dispersions of the two subclusters  are $1130^{+120}_{-320}$km\,s$^{-1}$ for the A2146-A subcluster and $760^{+360}_{-340}$ km\,s$^{-1}$ for the A2146-B subcluster, with a $68\%$ confidence interval (CI). It is important to note that the distribution of radial velocities is under-sampled and given that A2146 has recently undergone a merger, the distributions are possibly non-Gaussian. So the calculated velocity dispersions characterise the distribution, with the caveat that the sample may be non-Gaussian.

Application of the biweight estimator and bootstrap error approximations were carried out using the astropy library in the Python programming language\footnote{(http://docs.astropy.org/en/stable/)}. In a bootstrap technique, a $68\%$ CI is found for the MAD. This approach draws many samples with replacement from the data available, estimates the MAD from each sample, then rank orders the means to estimate the 16 and 84 percentile values for $68\%$ CI. Unlike assuming normal distributions to calculate $68\%$ CI, the results calculated from the bootstrap are robust and can compensate for a data distribution that is far from normal.

The velocity distribution of cluster members is shown in the histogram in Fig.\,4. The cluster members in A2146-A are shown in blue, and the cluster members in A2146-B are shown in red. Gaussian profiles for each subcluster are overlaid and centered at the mean determined from a biweight estimator \citep{beers2}.

\section{A2146 Dynamics}
\subsection{Virial Mass}
A common way to estimate the mass of a galaxy cluster is through the virial theorem, which assumes dynamical stability (e.g. \citealt{araya}), and therefore relates the time averaged kinetic energy to the potential energy  as
\begin{equation}
\langle T\rangle = -\frac{1}{2} \sum_{i}^{N} V_i\,.
\end{equation}
We calculate the virial masses below and tabulate them in Table \ref{table:nonlin}, along with the SZ and X-ray derived masses from the literature (see Section 6). However, we caution that the Abell 2146 system is not dynamically relaxed and discuss the potential biases of the mass measurements in Section \ref{discussion}. The virial theorem relates the gravitational potential energy to the kinetic energy (relative velocities) of the cluster members and the mass is calculated via the following relation
\begin{equation}
M_{vir} = \frac{3\pi}{G}\frac{\sigma_v^{2}}{R_{H}},
\end{equation}
where $\sigma_{v}$ is the velocity dispersion calculated in Section \ref{velocity}. $R_{H}$ is the mean harmonic radius given by
\begin{equation}
\frac{1}{R_{H}} = \frac{2(N-2)!}{N!} \sum_{i<j} \frac{1}{\|r_{ij}\|}\,,
\end{equation}
where $r_{ij}$ is the separation vector between the $i^{th}$ and $j^{th}$ galaxies and $N$ is the total number of objects. The benefit of using a mean harmonic radius is that it more accurately measures the effective radius of the gravitational potential of the cluster members  \citep{araya}, important when considering the virial state of a system. This radius also has the benefit of being independent of the cluster centre allowing it to reflect internal structure of the system, putting extra weight on objects that are closer together \citep{araya}. The virial radius for the entire system is related to the mean harmonic radius by $r_{\rm vir}\approx 2R_{H}=0.64$\,Mpc (e.g. \citealt{girardi}; \citealt{merchan}). 

Taking the values calculated for the velocity dispersions of each of the subclusters along with the harmonic radius for each, $68\%$ CI virial mass estimates are found to be $M_{\rm vir}= 6.0^{+1.3}_{-2.9} \times 10^{14}$  M$_{\odot}$ for A2146-A and $M_{\rm vir}= 2.5^{+3.0}_{-1.7} \times 10^{14}$ M$_{\odot}$ for A2146-B (we note here that subcluster B was found not to be well described by a Gaussian distribution), giving a mass ratio of 2.4. As an estimate for the total mass, we take the sum of the two estimates, since the assumption that the individual clusters are virialized is likely more appropriate than assuming that the system as a whole is virialized. This yields $M_{\rm vir}= 8.5^{+4.3}_{-4.7}\times 10^{14}$M$_{\odot}$. Note that the error bars do not reflect systematic errors which can change the mass estimate, even when the system is not in the throws of a merger; this simple dynamical estimate assumes a virialized spherical mass distribution with isotropic velocity dispersion.

The large errors in the mass estimates are due to the large errors inherent to the velocity dispersions. Since the virial mass is proportional to the square of the velocity dispersion, the error on the mass estimates is quite large. The $68\%$ CI level corresponds to a $1\sigma$ uncertainty. For a discussion of the impact of systematic and model dependent errors on the mass estimates, see Section \ref{discussion}.

Since the estimates of mass from the kinematics of galaxies make several simplifying assumptions, we consider now an alternative estimate of the mass of each subcluster derived from a scaling relation between velocity dispersion and mass. \citet{EV} used N-body simulations to obtain a prediction for the relationship between mass and velocity dispersion of dark matter halos. The advantage of this type of approach is that it includes more complex dynamical behaviour. Although their simulations were carried out using only dark matter, since galaxies and dark matter particles can both be treated as collisionless particles moving in a cluster potential, they are expected to have approximately the same velocity dispersion (\citealt{OK}; see also discussion in \citealt{EV}). Using the \citet{EV} relation 
\begin{equation}
M_{200}E(z) = 9.358\times 10^{14} \times (\sigma_{v}/1000)^{2.975} h^{-1}M_{\odot}\,,
\end{equation}
with the parameter $E(z)=1.264$ at the cluster redshift and for the cosmology we work in, and determining the mass estimate for each cluster gives a total mass of $M_{200} = 2.0^{+2.0}_{-1.4} \times 10^{15} M_{\odot}$.

The observed non-Gaussianity of A2146-B in conjunction with the system not being dynamically relaxed will contribute to an overestimation of the velocity dispersion for each subcluster component. The \citet{EV} scaling relation does not account for these and therefore it is likely that the mass will be over estimated as well.

Specifically for merging clusters, there have been studies using simulations to investigate the accuracy of various ways of estimating mass \citep{pinkney, zu} or focusing on specific merger systems (e.g. \citealt{zu}). For clusters with a mass ratio of 3:1, \citet{pinkney} show that the mass inferred from a simple application of the virial theorem, and when sub-clustering is ignored in the analysis, can be biased by as high as a factor of two. The maximum bias arises when the clusters are merging with an axis along the line of sight, and is somewhat less in mergers with an axis close to the plane of the sky, though still up to a factor of 1.5. The hydrodynamic simulations of \citet{zu} are consistent with the earlier work of \citet{pinkney}. \citet{zu} also support accounting for separate subclusters when determining mass estimates from the velocities of galaxies in a merger system.

\begin{figure}
\includegraphics[width=0.48\textwidth]{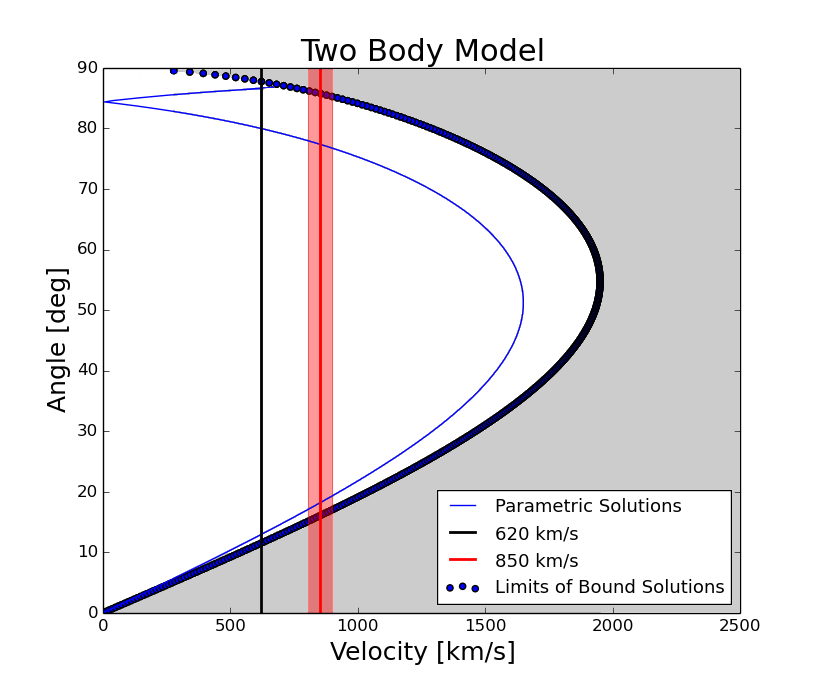}
 \caption{Projected angle from the plane of the sky, $\alpha$, plotted against the difference in radial velocities of the the subclusters, $V_{R}$. The black vertical line gives the solutions for $V_{R}= 620$\,km\,s$^{-1}$ and the red for $V_{R}= 850 \pm 60$\,km\,s$^{-1}$. The red shaded area around $V_{R}$ represents the errors in radial velocity from the cross-correlation. The dotted curve represents the limits of the bound solutions as calculated from Eq.\,8. The gray region represents the area where only unbound solutions would exist. The intercept of the vertical lines gives solutions corresponding to $\alpha = 13^{\circ} - 19^{\circ}$, $V=2600 - 2800$\,km\,s$^{-1}$, and $R=0.70 - 0.72$\,Mpc. \label{two}}
\end{figure}

\subsection{Two-Body Dynamics}
\label{twobody}
We undertake a dynamical analysis of the system using a two-body model, making the assumption that our system consists of two dominant subclusters, which is reasonable based on the projected density map (see Fig.\,\ref{density}). We use the methods adopted by Beers et al  \citep{beers1}, to which the reader is referred for further details, and also follow the formalism presented in e.g. \citet{girardi}, \citet{barrena} and \citet{ferrari}. Taking the total mass of the system derived from the virial mass estimates and the difference in radial velocities of the two subclusters, we calculated the angle between the merger axis and the plane of the sky, and the age of the merger. Note that this method assumes point masses on radial orbits with no net rotation. The subcluster halos will also most likely be overlapping in a recent merger. Making these assumptions in this analysis will add systematic and model dependent uncertainties to the results obtained. Therefore the solutions will be approximations to the dynamics of A2146. The coupled equations, which describe the time evolution of the system, are closed by setting $t=t_{0}=3.4 \times 10^{17}$ s, the age of the universe at the Abell 2146 system redshift $z=0.2323$. 

The parametric solutions to the equations of motion are given as follows:
\begin{equation}\label{it1}
R=\frac{R_{M}}{2}(1-\cos(\chi))\,; 
\end{equation}
\begin{equation}\label{it2}
t= \Big(\frac{R_{M}^{3}}{8GM}\Big)^{1/2} (\chi - \sin(\chi))\,:
\end{equation}
\begin{equation}\label{it3}
V=\Big(\frac{2GM}{R_{M}}\Big)^{1/2} \frac{\sin(\chi)}{1-\cos(\chi)}\,,
\end{equation}
where $R$ is the spatial separation, $V$ is the relative velocity between the two BCGs, $R_{M}$ is the separation of the two subclusters at maximum expansion, $M$ is the total virial mass of the system, and $\chi$ is the so-called developmental angle.

The limits of the bound solutions can be found by considering the gravitational binding energy in the following relation
\begin{equation}
V_{R}^{2} R_{P} \le 2 G M_{Total}\sin^{2}(\alpha)\cos(\alpha)
\end{equation}
where $V_{R}$ is the difference in radial velocities of the two subclusters and $R_{P}$ is the projected separation of the cores of the two subclusters.

Since the velocity distributions of the subclusters, A2146-A and A2146-B, do not precisely follow a Gaussian distribution, the difference in the mean radial velocity of the subclusters from a Gaussian fit (from Fig. \ref{red_dist}) will not yield an accurate value of $V_{R}$. We therefore opt to use the difference in radial velocities of the two BCGs in each of the subclusters as a determination of $V_{R}$. 

\begin{figure}
\includegraphics[width=0.48\textwidth]{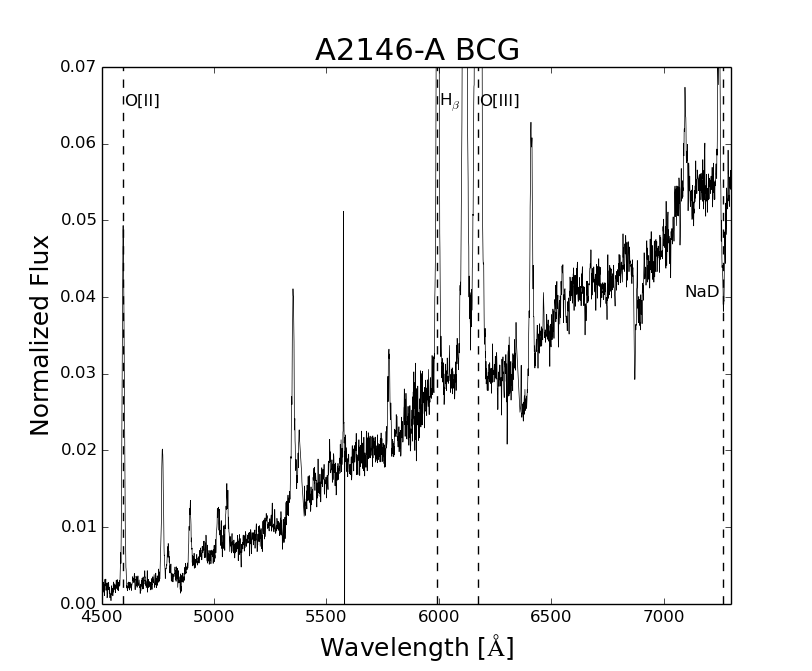}
 \caption{Zoomed in view of the spectrum for the BCG in A2146-A.  Both absorption and emission features are clearly visible and the radial velocity of the emission features are offset by $\approx +230$\,km\,s$^{-1}$ from the BCG. Both absorption and emission features are labeled. \label{zoom}}
\end{figure}

The spectra of the two BCGs are shown in Section 3 (Fig.\ref{BCG_b}). The BCG in A2146-B has a strong absorption spectrum with well defined features, as is typical of BCGs. The spectrum of the BCG in A2146-A, however, shows many emission lines. The three most prominent features are $H_{\beta}$ and two [OIII] lines. Zooming in on this spectrum (Fig.\ref{zoom}) shows the characteristic absorption spectrum expected for an elliptical galaxy with additional strong emission features. 

While most of the absorption features are blended with emission features, making it difficult to accurately determine redshift, the Na D (rest wavelength of 5895 \AA) feature is relatively unblended and yields a velocity of $\sim$69,612 km s$^{-1}$. This is a lower redshift than determined through cross correlation methods for the BCG which rely on the emission lines (69,897 km s$^{-1}$). \citet{canning} show that much of the emission line gas in the A2146-A BCG is offset in a plume from the centre of the galaxy which may be the cause of the discrepancy in the radial velocities of the emission and absorption lines.

Calculating $V_{R}$ as the difference between the absorption line spectrum of A2146-B and the emission spectrum of A2146-A gives a value of $V_{R}=850 \pm 60$ km\,s$^{-1}$. Using instead the Na D absorption feature yields a value of $V_{R}=620$ km\,s$^{-1}$. Nominally, it would be advantageous to take the velocities of the absorption lines of the galaxies, however, as the BCG in subcluster A is also offset from its predicted location - leading, rather than trailing, the cool core, which may indicate that the BCG is not in this case a good tracer of the center of the mass distribution - we opt to consider the uncertainty as the relatively broad range in velocities from the emission and absorption spectra.

We therefore take $V_{R}$ to be in the range $620$\,km\,s$^{-1}$ to $850 \pm 60$\,km\,s$^{-1}$ in the two-body analysis. $R_{P}$ is taken as the projected separation of the two BCGS. At the cluster redshift this corresponds to 0.69 Mpc.

The radial velocity and projected radius are related to the system parameters as follows:
\begin{equation}
V_{R}= V sin(\alpha) ;  R_{P}=Rcos(\alpha)\,.
\end{equation}

Solving equations \ref{it1}, \ref{it2}, and \ref{it3} iteratively yields the solutions shown in Fig.\,6. Plotting $V_{R}$ shows the solutions for our system. While there are solutions of $\alpha$ close to $80^{\circ}$, the presence of the observed shocks in the X-ray analysis  \citep{russell10} and the estimated value of $17^{\circ}$  \citep{canning}, suggests that the merger is close to the plane of the sky. This allows us to take the range of solutions for $\alpha = 13^{\circ}$ to $19^{\circ}$, which are very close to the values derived by \citet{canning}. The results from the two-body model analysis are $V=2600 - 2800$\,km\,s$^{-1}$ for the relative velocity, and $R=0.70 - 0.72$\,Mpc for the separation. Assuming a collision site of $\approx{R/2}$,  half way between the two subclusters, yields a merger age of $0.24- 0.28$ Gyr. 

\section{Discussion}
\label{discussion}

X-ray observations of Abell 2146 have revealed the system has two clear shock fronts and a gas structure similar to the Bullet cluster \citep{markevitch02} showing it must be in the throes of a major merger. The detection of two clear shock fronts also indicate that the merger should be in an early stage. However, the Abell 2146 system presents a couple of challenges to that hypothesis. 

The first is that the A2146-A subcluster BCG is located behind the bow shock and in the wake of the ram pressure stripped cool core rather than leading it (see \citealt{russell10} and \citealt{canning}). As mentioned in Section \ref{intro} shortly after a galaxy cluster collision the intra-cluster medium is expected to lag behind the bulk of the mass distribution traced by the dark matter and the major concentrations of cluster galaxies (e.g. \citealt{clowe}). 

One possibility for the unusual location of the BCG could be that the cluster merger is not in its early stages (e.g. \citealt{hallman}). In the later stages of a galaxy cluster merger the collisionless matter distributions slow due to gravity allowing the gas to catch up and be accelerated in a `gravitational sling shot' beyond the dominant mass distribution. Other possibilities are that the system may not be adequately represented by a two-body collision with the cluster containing complex substructure, or that the merger is somewhat off-axis.

Examining the redshifts of our target galaxies we find 63 galaxies are likely members of the A2146 system. An analysis of the projected density peaks of the galaxies (see Section \ref{subclustering}) shows that the Abell 2146 system is dominated by two subclusters. Our data do not provide evidence for other more complicated structures such as other merging groups or subclusters. We use the KMM algorithm to assign subcluster membership to the two subclusters and find 35 subcluster member galaxies in A2146-A and 28 in A2146-B. Application of the DS test to determine substructure was inconclusive. However, we emphasise  that in \citet{DS} and in \citet{pinkney}, the sensitivity of the DS test to substructure in merger systems is shown to be diminished when the subclusters are of similar mass and when their merger axis is close to the plane of the sky, as is likely the case in Abell 2146.   

Given that the system appears to be fairly well described by two dominant subclusters we undertake a simple two-body analysis of the subcluster velocities and positions (section \ref{twobody} and Fig.\,\ref{two}) to determine the properties of the merger. As the velocity distributions of the subclusters are not both well described by Gaussian distributions we choose to use the BCGs as a tracer of the centre of the subcluster mass distributions. We note here that in the case of subcluster A2146-A this may be incorrect since, as mentioned above, the BCG is offset from the expected location of the centre of the mass distribution; whilst the centroid of the distribution of red sequence galaxies is approximately coincident with the X-ray cool core, the BCG appears to lag behind the cool core. If a collision has offset the BCG we may therefore expect the BCG velocity not to accurately trace the subcluster velocity centroid. The A2146-A BCG not only shows absorption features but also strong emission features in the spectrum which have a velocity offset of $\sim$230 km~s$^{-1}$. This is significantly larger than the errors in the velocities from cross-correlation ($\sim$30 km~s$^{-1}$). The centroid of the subcluster velocity distribution falls in between the emission and absorption values (A2146-A; BCG emission 69,897 km s$^{-1}$, BCG absorption 69,612 km s$^{-1}$, Centroid (KMM) 69,878 km s$^{-1}$: A2146-B; BCG absorption 68,827 km s$^{-1}$, Centroid (KMM) 69,710 km s$^{-1}$; WYFFOS spectra published in \citealt{canning} determined similar velocities of 69,753 and 68,940~km~s$^{-1}$, from single Gaussian fits to the emission and absorption spectra, for BCG A and B respectively). We therefore choose to use the full range of velocities from the absorption and emission lines in the A2146-A BCG in the two body analysis to allow  a conservative estimate of the merger axis angle and timescale.

Despite these uncertainties in cluster velocity centroid, the merger age, for bound solutions close to the plane of the sky, is still predicted to be young (0.24$-$0.28\,Gyr), similar to that determined for the Bullet cluster \citep{barrena}. The subclusters are therefore likely being observed shortly after first core passage. The unusual BCG position is therefore unlikely to be due to a gravitational `slingshot' of the gas past the galaxies (see e.g. \citealt{hallman}).  

The merger axis is determined to be very close to the plane of the sky with $\alpha = 13^{\circ}$ to $19^{\circ}$. The additional $\alpha$ solution (top of Fig. \ref{two}) that would correspond to a line-of-sight merger is unlikely due to \citet{russell10, russell12} observing both forward and reverse X-ray shock fronts. Observations of merger shocks are rare in X-ray observations as they are easily diluted by projection effects, they are therefore most prevalent in systems which are close to the plane of the sky. We therefore conclude that the unusual BCG location is not due to projection effects in a line-of-sight merger. 

These findings for the merger timescale and merger axis are also consistent with the presence of the two large shock fronts seen in {\it Chandra} X-ray observations \citep{russell12}, since later in the merger the shock fronts would not be visible, and for a larger angle with respect to the plane of the sky their sharp edges would be smeared by projection effects.

Although the two-body model is useful for estimating the dynamical history of merger systems, the assumptions that the clusters are point masses with constant mass over time, start to break down when they begin to merge. The analysis also does not include a change in relative velocity of the clusters with time.
In the context of the Bullet Cluster, \citet{nusser} discusses several limitations of the two-bodel model, for example how dynamical friction and tidal stripping are not included in this type of analysis. Both of these effects act to reduce the relative speeds of the cluster components. Although hydrodynamic simulations are the most accurate way to model the dynamics of merger systems, there are prohibitive in terms of the complexity of setting up the simulations, and the computational expense. \citet{dawson} presents a novel approach where the computational requirements fall somewhere between the two-body model and hydrodynamic simulations. By comparison with full hydrodynamic simulations of the Bullet Cluster, this method is accurate to about 10\% in determining parameters describing the clusters and their merger trajectory. Required inputs to this dynamical modeling approach include spectroscopic redshifts for cluster members, arising from this work, and additional mass estimates available from the weak lensing analysis (King et al. 2015). Thus application of the \citet{dawson} method to Abell 2146 will be the topic of future work.

\begin{figure}
 \centering
\includegraphics[width=0.48\textwidth]{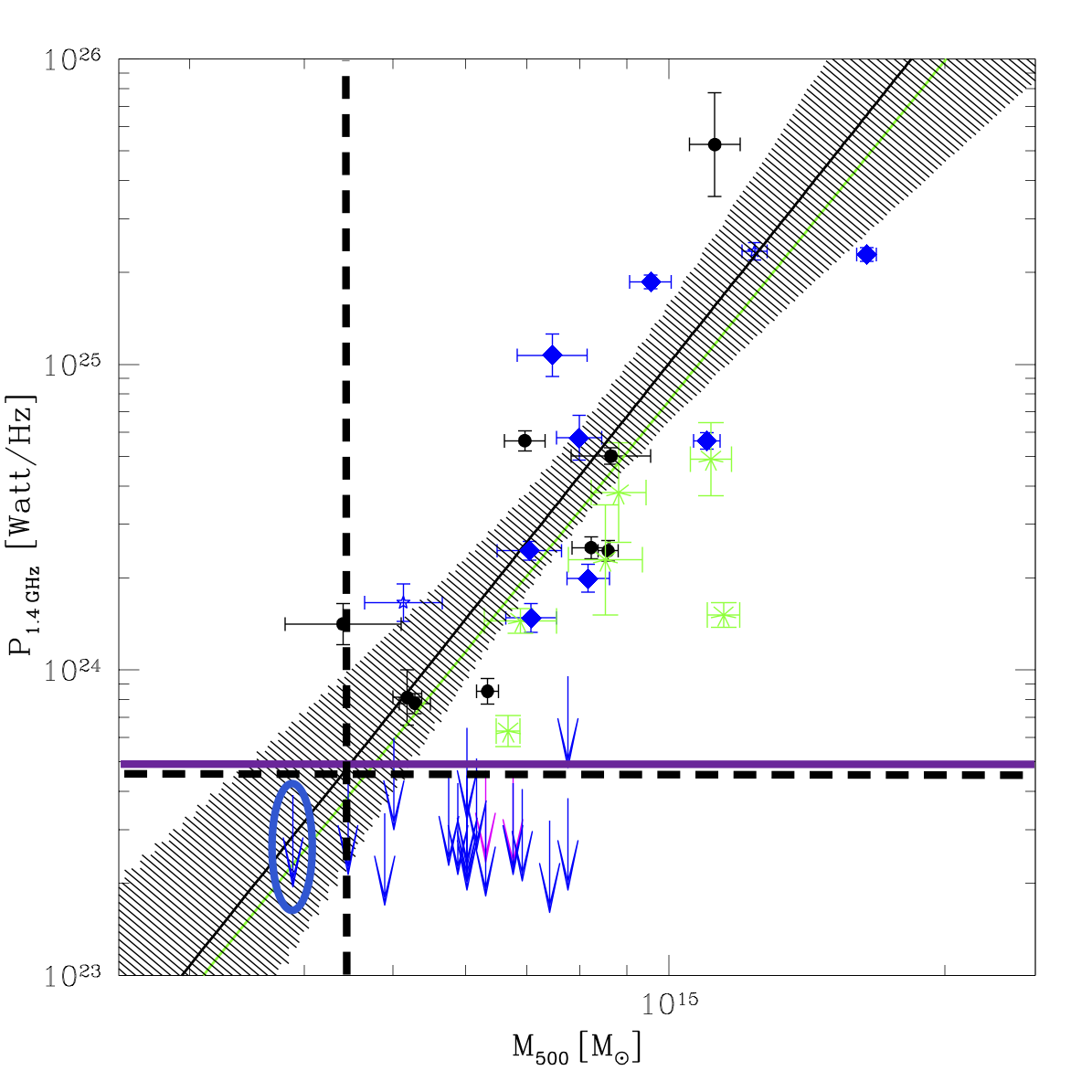}
 \caption{ Distribution of clusters in the $P_{1.4} - M_{500}$ diagram (Fig. 3 from Cassano et al. (2013)). The blue circles were changed to diamonds for clarity and additional lines are added corresponding to dynamical and X-ray analysis. The various different symbols indicate halos belonging to the extended Giant Metrewave Radio Telescope (GMRT) Radio Halo Survey \citep{cassano}, or EGRHS, (blue diamonds); halos from the literature (black circles); halos with very steep spectra (USSRH, green asterisks); A1995 and Bullet cluster (blue stars); cool core clusters belonging to the EGRHS (magenta arrows). Best-fit relations to giant RHs only (black lines) and to all RHs (including USSRH, green dashed lines) are reported. The circled blue arrow is A2146 with an $M_{500}$ and $P_{1.4}$ reported by Cassano et al. (2013). The dashed line shows A2146 and it's derived $M_{500}$ from dynamical analysis. This corresponds to a radio halo power upper limit of $P_{1.4} \approx 4\times 10^{23}$W Hz$^{-1}$ for a mass of $M_{500} =  4.3 \times 10^{14} M_{\odot}$. The horizontal purple line marks the calculated $3\sigma$ upper limit for Abell 2146 from \citet{russell11}. \label{RH}}
\end{figure}

The second challenge for Abell 2146 is its apparent lack of radio emission, neither radio relics, which would be expected to be associated with the shock fronts, nor a radio halo have been seen \citep{russell11}. Large-scale diffuse radio emission in clusters has only been found in merging systems and as such it could be a useful diagnostic of the statistics of dynamically relaxed and merging systems. However, the mechanisms which form this diffuse radio emission and the conditions required for their formation are not yet known.

Diffuse radio emission on Mpc-scales in clusters requires local particle acceleration where either a pre-existing electron population is re-accelerated by merger turbulence (e.g. \citealt{brunetti01, petrosian}) or emitting electrons are continuously injected by collisions between cosmic ray protons and thermal protons (e.g. \citealt{dennison}; for reviews of extended radio emission in clusters see \citealt{feretti, ferrari2008, feretti12}).  The recent discovery of extremely steep radio halo spectra (e.g. \citealt{brunetti, dallacasa}) together with the non-detection of nearby clusters at $\gamma$-ray wavelengths (e.g. \citealt{brunetti12, ackerman}) favours turbulent re-acceleration models over the `secondary' continuous injection models.

Correlations between the synchrotron power in radio halos, $P_{\mathrm{radio}}$, and the galaxy cluster X-ray luminosity, $L_{\mathrm{X}}$ and temperature, $T_{\mathrm{X}}$, have been found, which also show a bi-modailty for merging and non-merging galaxy clusters (e.g. \citealt{liang, kempner, bacchi, cassano06, brunetti7}). Clusters with detected radio halos have $P_{\mathrm{radio}}$ correlated with $L_{\mathrm{X}}$ and show a disturbed morphology consistent with merging where there is no radio halo detected, and with only upper limits on radio luminosity, the clusters generally appear more relaxed and lie roughly an order of magnitude below this correlation \citep{brunetti, rossetti}. However, some clusters, such as Abell 2146, have been found which confound this bi-modality in the X-ray  (e.g. \citealt{russell11, giovannini, cassano10}). 

\cite{russell11} showed the 3$\sigma$ radio halo upper limit is a factor $\sim$5 times lower than predicted given its X-ray luminosity. X-ray luminosity is boosted in a merger and this factor is greater for young mergers \citep{ricker} so the confirmed young age of the Abell 2146 system (being only $0.24-0.28$ Gyr old) and the near equal masses of the subclusters indicates its X-ray luminosity is likely to be enhanced by a factor of $\sim$2-5. A boost this significant could place Abell 2146 back on the relation for merging clusters. However, the age since core passage is similar to other systems in which radio halos have been detected.

\cite{brunetti12} examine a sample of radio halo clusters from the GMRT Radio Halo Survey and find that cluster synchrotron emission is suppressed on timescales of only a few hundred Myrs. The simulations of \citet{donner}, albeit for a 1:7 mass ratio merger, suggest that the radio halo luminosity should be increasing rapidly during infall and through core passage, with the X-ray and radio luminosity, within a few hundred Myr, being boosted by factors of 2 and 30, respectively. They find the radio emission starts to decay roughly 0.5 Gyr after core passage. While timescale may be relevant in boosting the X-ray luminosity and therefore in where Abell 2146 should sit on the $P_{\mathrm{radio}}$-$L_{\mathrm{X}}$ relation, these and other simulations suggest that the timescale since core passage for Abell 2146 is not too short or too long for the generation of a radio halo.

Cluster mass is thought to be a key component of radio halo generation in galaxy clusters \citep{cassanobrunetti}. \cite{basu} examine the radio-SZ correlation for clusters on the basis that the integrated SZ signal is a more robust indicator of the cluster mass regardless of dynamical state than the X-ray luminosity (e.g. \citealt{motl, nagai}). A correlation is found for the radio power and integrated SZ signal, $Y_{\mathrm{SZ}}$ for the sample of clusters and no bi-modality is detected between those clusters which are relaxed and those which are disturbed. The authors suggest the bi-modality with X-ray properties may be due to a bias in the radio quiet, X-ray selected clusters, being selected towards lower mass cool core clusters. However, they note that their sample size for radio quiet systems is small and so a weaker bi-modality might still exist. \citep{cassano} find the converse, that a bi-modailty exists, in both the $P_{\mathrm{radio}}$-$L_{\mathrm{X}}$ and $P_{\mathrm{radio}}$-$Y_{\mathrm{SZ}}$ relations with all relaxed systems having radio halo upper limits which sit $>$2$\sigma$ below the relation. However, they suggest mass is also important and that the bi-modality only exists for `SZ-luminous' $Y_{\mathrm{SZ}}>6\times10^{-5}$~Mpc$^{2}$, or alternatively massive, $M_{500}>5.5\times10^{14}$~M$_{\odot}$ galaxy clusters. 

The scaling relation of \citet{EV}, while robust, does not account for the observed subclustering or dynamical state of Abell 2146. While our individual subcluster masses similarly suffer from the dynamics of the merger the total mass as the sum of the subcluster masses rather than assuming one virialised system should yield value more in line with previous estimates. We therefore use a total mass of $M_{\rm vir}= 8.5^{+4.3}_{-4.7} \times 10^{14}$ M$_{\odot}$ giving a mass ratio of approximately 2.4:1 (A2146-A to A2146-B). The large uncertainties in the mass estimates preclude determining which subcluster is in fact the more massive component with a high degree of confidence. However, the system appears to have undergone a recent relatively equal-mass merger. The X-ray analysis found that the bow shock and upstream shock are of comparable strength, which is consistent with a close to equal mass merger, however a 1:4 mass ratio with A2146-B being more massive than A2146-A was favored \cite{russell10}. For higher mass ratio systems, such as the Bullet cluster  \citep{clowe} with a ratio of $\approx$10:1, the perturbation to the gravitational potential of the merger is much smaller and therefore will produce a much weaker upstream shock.  Our dynamical mass ratio estimate is consistent with a weak lensing analysis of the system (King et al. 2015).

The NFW (Navarro, Frenk \& White 1996) $r_{200}$ value of the system is approximately equal to the virial radius, or to twice the harmonic radius calculated in Section 5.1, yielding a value of $r_{200} = 0.64$\,Mpc. Assuming an NFW mass profile and a halo concentration of 4 typical of massive galaxy clusters (see collation from King \& Mead 2011), an estimate of $r_{500} \approx 0.42$\,Mpc was calculated, corresponding to $ M_{500}=5.4^{+2.7}_{-3.0} \times 10^{14}$  M$_{\odot}$. 

To take into account the overestimate in mass due to the system having undergone a recent merger, a surface pressure term (SPT) can be subtracted from the viral mass to get a more accurate estimate of the true mass of the system \citep{the}. Assuming the velocities are isotropic, the SPT correction can be taken as $SPT=0.2 \times M_{\rm vir}$  \citep{girardi}. The reduction of the mass due to the SPT correction would result in a lower mass estimate of $M_{200}=6.8 \times 10^{14}$ M$_{\odot}$ and $M_{500}= 4.3 \times 10^{14}$ M$_{\odot}$.

For Abell 2146, SZ observations  \citep{rodriguez} yielded a total system mass inside $r_{200}$ of $M_{200,\rm SZ} = 4.1 \pm 0.5 \times 10^{14} h^{-1}$ M$_{\odot}$ (or $5.6^{+0.7}_{-0.7} \times 10^{14}$ M$_{\odot}$ using the same cosmology as this paper i.e. $h= \rm H_{0}/70$ km\,s$^{-1}$Mpc$^{-1}$). This is a factor of $\sim$1.5 smaller than our dynamical virial mass estimate for the total system mass not including a SPT correction, or roughly 1.2 times smaller with the SPT correction, bringing the calculated $M_{200}$ within the errors of the SZ mass estimate. The $M_{500}$ and $M_{200}$ values calculated from the SZ (taking $h=H_{0}/70$ km\,s$^{-1}$Mpc$^{-1}$), our dynamics analysis, and the X-ray analysis of \cite{russell12} are compiled in Table 1.

When correcting $M_{500}$ for the SPT, and using the direct scaling of $M_{500}-P_{1.4}$,  shown in Fig.\,7 (Fig.\,3 from \citealt{cassano}), we find an expected radio halo power of $P_{1.4} \approx 4\times 10^{23}$ Watt Hz$^{-1}$, marked with the dashed black horizontal line in Fig.\,7. The solid purple horizontal line marks the calculated $3\sigma$ upper limit for A2146 from the GMRT radio observations by \citet{russell11}. This observational upper limit is therefore very similar to the radio power predicted by the observed $M_{500}-P_{1.4}$ correlation for merging systems, using a dynamical mass estimate. 

A lower $M_{500}$ value could be the leading factor as to why no extended radio emission is detected, and this is accommodated within the errors on the dynamical mass, particularly if we include the SPT. As the radio observing sensitivity increases in the future, faint extended radio emission may well be detected and the possibility of a mass threshold for radio halos in merging clusters will be clarified. Furthermore, the prospects for assessing the dependence of radio halo production on other factors will improve with studies of large samples of major cluster mergers.  

\begin{table}
\caption{Summary of Mass Estimates for Abell 2146. All masses are in units of M$_{\odot}$.} 
\centering 
\begin{tabular}{c | c | c} 
\hline\hline 
Mass & $M_{500}$ & $M_{200}$  \\   
\hline 
Dynamics &  $5.4^{+2.7}_{-3.0} \times 10^{14}$ &  $8.5^{+4.3}_{-4.7} \times 10^{14}$ \\
SZ  & $4.3^{+0.6}_{-0.6}  \times 10^{14}$ &  $5.6^{+0.7}_{-0.7} \times 10^{14}$ \\
X-ray & $\approx 7 ^{+2}_{-2} \times 10^{14}$ & ------ \\ [2ex]
\hline 
\end{tabular} 
\label{table:nonlin} 
\end{table}

\section{Conclusions}
We present a dynamical analysis of the galaxy cluster merger system Abell 2146 using multi-object spectroscopic observations from the {\it Gemini} North Telescope. We employ substructure tests such as the Dressler-Schectman test \citep{DS}, analysis of an adaptive kernel density map \citep{pisani}, and application of the KMM algorithm \citep{ashman} using the position information of the galaxies to conclude that the system is dominated by two main substructures, and is likely well described by a two-body collision. One cluster is located in the south-east (A2146-A) and the other in the north-west (A2146-B). An average redshift of $z=0.2323$ was calculated with velocity dispersions of $1130^{+120}_{-320}$km\,s$^{-1}$ and $760^{+360}_{-340}$ km\,s$^{-1}$ respectively for the A2146-A and A2146-B  subclusters. The main results are:

\begin{enumerate}
\item From the 63 spectroscopically confirmed A2146 system members, we estimate a total virial mass of $M_{\rm vir}= 8.5^{+4.3}_{-4.7} \times 10^{14}$ M$_{\odot}$, without any correction applied for the $\sim$20 per cent boost in mass due to the system having undergone a recent merger (e.g. \citealt{the}).
\item The two subclusters have a close to 2.4:1 mass ratio indicating that this is a major merger. 
\item A two-body analysis of the system yielded an angle between the merger axis and plane of sky of approximately $13^{\circ}-19^{\circ}$ and a time since core passage of the merger of approximately $0.24-0.28$ Gyr. 
\end{enumerate}

These findings for the dynamics are consistent with the observation of large shock fronts on {\it Chandra} X-ray images (Russell et al. 2010; 2012). We conclude from our analysis that the unusual location of the BCG, behind, not in front of the shock front and X-ray cool core, is not due to the merger being in a late stage of its evolution with the gas experiencing a gravitational sling-shot beyond the main mass distribution.

Despite the lack of an observed radio halo, simulations of our calculated timescale and near equal mass ratio of the merger would seem to favor the generation of a radio halo (e.g. \citealt{donner, brunetti12}). However, the final mass of the system, and therefore necessarily the progenitor masses of the two subclusters, is fairly low, compared with other similar cluster mergers. \cite{cassano} suggest that a bi-modality in radio halo properties for merging and non-merging systems may require a total mass threshold of about $M_{500}>5.5\times10^{14}$~M$_{\odot}$. Uncertainties on the mass estimates are very large, however, Abell 2146 would lie close to this threshold. Deeper radio data on this and other merging systems that appear to lack radio halos would be critical for testing whether a threshold in mass exists. Magnetohydrodynamic simulations of the system, constrained by the wealth of observational data, will further our understanding of the distribution of luminous and dark matter in the system, as well as the radio emission. For the Bullet Cluster, recent work by \citet{lage} illustrates that this approach provides much insight into merger systems. Application of the \citet{dawson} method for dynamical modeling will provide a more accurate description of the geometry of the merger than provided by the two body model, also greatly narrowing down the parameter space for MHD simulations.

Conflicting scenarios have been proposed where merger events may be a catalyst to spur star formation -  for example shock fronts moving through the ICM and triggering further starbursts \citep{mihos, bekki} - or to quench star formation via ram-pressure stripping (Fujita et al. 1999).  Detailed analysis of the [O {\sc ii}], [O {\sc iii}], $H_{\alpha}$, and $H_{\beta}$ spectral lines is currently underway (Lee et al. in prep), in order to estimate the star formation rates in the cluster galaxies as discussed in \citet{ferrari}. 

\section{Acknowledgements}
JAW acknowledges support from NASA through the Columbia Crew Memorial Undergraduate Scholarship of the Texas Space Grant Consortium. REAC acknowledges support from a scholarship from the Cambridge Philosophical Society and a Royal Astronomical Society grant. JAW, LJK, BEL and JEC would like to acknowledge support from The University of Texas at Dallas, and JEC acknowledges NASA for a Graduate Fellowship of the Texas Space Grant Consortium.
ACE acknowledges support from STFC grant ST/I001573/1. HRR and ACF acknowledge support from ERC Advanced Grant Feedback. JAW and LJK thank Mike Kesden for discussions on merger kinematics and Ylva Schuberth for discussion on data reduction and IRAF. We would all like to thank Cassano et al. (2013) for kindly allowing the use of their plot used in Fig. 8. We thank the referee for detailed comments which greatly improved the paper. We would also like to thank Jessica Mink for very helpful advice on the SAO templates for use in XCSAO.
This research has made use of the NASA/IPAC Extragalactic Database (NED) which is operated by the Jet Propulsion Laboratory, California Institute of Technology, under contract with the National Aeronautics and Space Administration.

Based on observations made with the NASA/ESA Hubble Space Telescope, obtained through program 12871 through the Space Telescope Science Institute, which is operated by the Association of Universities for Research in Astronomy, Inc., under NASA contract NAS 5-26555. 

Based on observations obtained at the Gemini Observatory, which is operated by the 
Association of Universities for Research in Astronomy, Inc., under a cooperative agreement 
with the NSF on behalf of the Gemini partnership: the National Science Foundation 
(United States), the National Research Council (Canada), CONICYT (Chile), the Australian 
Research Council (Australia), Minist\'{e}rio da Ci\^{e}ncia, Tecnologia e Inova\c{c}\~{a}o 
(Brazil) and Ministerio de Ciencia, Tecnolog\'{i}a e Innovaci\'{o}n Productiva (Argentina).

.

\appendix

\onecolumn
\begin{center}
\begin{longtable}{lllllllllll}
\caption[Cluster Members.] {Data for 63 spectroscopically confirmed cluster member galaxies in the field of Abell 2146. The radial velocities are calculated from RVSAO cross-correlation methods applied to our {\it Gemini} GMOS spectroscopic observations (PI Canning).} \label{long_tab} \\

\hline \hline \\[-2ex]
   \multicolumn{1}{c}{\textbf{ID}} &
   \multicolumn{1}{c}{\textbf{RA}} &
   \multicolumn{1}{c}{\textbf{Dec}} &
   \multicolumn{1}{c}{\textbf{$cz$}} &
   \multicolumn{1}{c}{\textbf{$\Delta cz$}} &
   \multicolumn{1}{c}{\textbf{Subcluster}} &
   \multicolumn{1}{c}{\textbf{Type}} &\\
   \multicolumn{1}{c}{\textbf{}} &
   \multicolumn{1}{c}{\textbf{(degrees)}} &
   \multicolumn{1}{c}{\textbf{(degrees)}} &
   \multicolumn{1}{c}{\textbf{(km~s$^{-1}$)}} &
   \multicolumn{1}{c}{\textbf{(km~s$^{-1}$)}} &
   \multicolumn{1}{c}{\textbf{(A/B)}} &
   \multicolumn{1}{c}{\textbf{(Emission/Absorption)}} &\\
[0.5ex] \hline
   \\[-1.8ex]
\endfirsthead

\multicolumn{11}{c}{{\tablename} \thetable{} -- Continued} \\[0.5ex]
  \hline \hline \\[-2ex]
   \multicolumn{1}{c}{\textbf{ID}} &
   \multicolumn{1}{c}{\textbf{RA}} &
   \multicolumn{1}{c}{\textbf{Dec}} &
   \multicolumn{1}{c}{\textbf{$cz$}} &
   \multicolumn{1}{c}{\textbf{$\Delta cz$}} &
   \multicolumn{1}{c}{\textbf{Subcluster}} &
   \multicolumn{1}{c}{\textbf{Type}} & \\
   \multicolumn{1}{c}{\textbf{}} &
   \multicolumn{1}{c}{\textbf{(degrees)}} &
   \multicolumn{1}{c}{\textbf{(degrees)}} &
   \multicolumn{1}{c}{\textbf{(km~s$^{-1}$)}} &
   \multicolumn{1}{c}{\textbf{(km~s$^{-1}$)}} &
   \multicolumn{1}{c}{\textbf{(A/B)}} &
   \multicolumn{1}{c}{\textbf{(Emission/Absorption)}} &\\
[0.5ex] \hline
  \\[-1.8ex]
\endhead

5 & 238.9647 & 66.37298 & 69484 & 24 & B & Ab \\
13 & 238.9836 & 66.36637 & 68014 & 130 & B & Ab \\
14 & 238.9839 & 66.37659 & 69128 & 47 & B & Ab \\
16 & 238.9883 & 66.38978 & 67482 & 35 & B & Ab \\
21 & 238.9973 & 66.37249 & 70186 & 16 & B & Ab \\
22 & 238.9987 & 66.36988 & 67705 & 10 & B & Em \\
23 & 239.0007 & 66.37329 & 68849 & 24 & B (BCG) & Ab \\
24 & 239.0027 & 66.36994 & 71123 & 37 & B & Ab \\
35 & 239.0064 & 66.36687 & 68517 & 17 & B & Ab \\
36 & 239.0085 & 66.3711 & 71163 & 31 & B & Ab \\
39 & 239.013 & 66.3703 & 71325 & 21 & B & Ab \\
43 & 239.0155 & 66.37942 & 68398 & 19 & B & Ab \\
47 & 239.0175 & 66.377 & 70649 & 20 & B & Ab \\
48 & 239.0196 & 66.35606 & 70406 & 31 & B & Ab \\
50 & 239.022 & 66.3901 & 69195 & 58 & B & Ab \\
54 & 239.0246 & 66.36791 & 71930 & 49 & B & Ab \\
55 & 239.025 & 66.3782 & 68802 & 21 & B & Em \\
57 & 239.0291 & 66.36815 & 71835 & 37 & B & Ab \\
60 & 239.031 & 66.3541 & 70504 & 78 & A & Ab \\
61 & 239.034 & 66.349 & 69249 & 78 & A & Ab \\
63 & 239.034 & 66.3764 & 69432 & 32 & B & Ab \\
67 & 239.0385 & 66.3488 & 71634 & 14 & A & Ab \\
68 & 239.0385 & 66.3294 & 68242 & 33 & A & Ab \\
71 & 239.0403 & 66.36081 & 72876 & 16 & A & Ab \\
76 & 239.0445 & 66.3462 & 71054 & 81 & A & Em \\
79 & 239.0495 & 66.36106 & 69924 & 27 & A & Ab \\
80 & 239.0505 & 66.3496 & 71365 & 25 & A & Ab \\
81 & 239.0505 & 66.3534 & 69371 & 43 & A & Ab \\
83 & 239.052 & 66.3567 & 72327 & 41 & A & Em \\
87 & 239.0535 & 66.3701 & 69925 & 38 & A & Ab \\
90 & 239.056 & 66.33381 & 68368 & 54 & A & Ab \\
92 & 239.0565 & 66.3565 & 67810 & 37 & A & Ab \\
94 & 239.058 & 66.3482 & 69897 & 26 & A (BCG) & Em \\
96 & 239.0601 & 66.34456 & 70033 & 22 & A & Ab \\
105 & 239.067 & 66.3461 & 69497 & 24 & A & Ab \\
108 & 239.0691 & 66.34259 & 68075 & 21 & A & Ab \\
109 & 239.0694 & 66.34006 & 71013 & 52 & A & Ab \\
110 & 239.07 & 66.3467 & 68929 & 16 & A & Ab \\
113 & 239.0762 & 66.3713 & 68790 & 28 & A & Ab \\
115 & 239.0768 & 66.36112 & 69552 & 31 & A & Ab \\
116 & 239.0795 & 66.3415 & 71610 & 18 & A & Ab \\
117 & 239.0799 & 66.34054 & 69059 & 32 & A & Ab \\
120 & 239.0827 & 66.34473 & 69121 & 39 & A & Ab \\
122 & 239.0835 & 66.3427 & 72489 & 42 & A & Em \\
123 & 239.0835 & 66.3564 & 71683 & 35 & A & Ab \\
128 & 239.0955 & 66.3482 & 70673 & 31 & A & Ab \\
135 & 239.1241 & 66.35526 & 68352 & 18 & A & Ab \\
138 & 238.989 & 66.3766 & 68624 & 30 & B & Ab \\
139 & 239.0091 & 66.36326 & 68960 & 22 & B & Ab \\
142 & 238.9803 & 66.36338 & 69323 & 23 & B & Ab \\
143 & 238.9758 & 66.36388 & 66669 & 25 & B & Ab \\
147 & 239.0616 & 66.33845 & 69860 & 14 & A & Ab \\
149 & 239.0802 & 66.33684 & 67965 & 25 & A & Ab \\
150 & 239.085 & 66.343 & 70321 & 16 & A & Ab \\
152 & 239.0406 & 66.34549 & 67638 & 18 & A & Ab \\
153 & 239.058 & 66.356 & 70031 & 22 & A & Ab \\
154 & 239.04 & 66.3502 & 70282 & 17 & A & Ab \\
156 & 239.0025 & 66.3748 & 69179 & 27 & B & Ab \\
157 & 239.016 & 66.3708 & 68956 & 27 & B & Ab \\
164 & 239.0625 & 66.3677 & 71160 & 39 & A & Em \\
165 & 239.0184 & 66.36444 & 69028 & 31 & B & Ab \\
168 & 239.0051 & 66.39083 & 70413 & 35 & B & Em \\
170 & 238.959 & 66.3771 & 69239 & 23 & B & Ab \\

\end{longtable}
\end{center}
\twocolumn

\label{lastpage}

\end{document}